\documentclass[11pt, a4paper]{article}
\usepackage[pdftex]{graphicx}
\usepackage{dcolumn}
\usepackage{bm}
\usepackage{amsmath}
\usepackage{amsfonts}
\usepackage{authblk}

\begin{document}

\title{Preconditioned dynamic mode decomposition and mode selection algorithms for large datasets using incremental proper orthogonal decomposition} 
\author[1]{Yuya Ohmichi}
\affil[1]{Aeronautical Technology Directorate, Japan Aerospace Exploration Agency, 7-44-1 Jindaijihigashi, Chofu, Tokyo, 182-8522, Japan}

\date{\today}

\maketitle

\begin{abstract}
This note proposes a simple and general framework of dynamic mode decomposition (DMD) and a mode selection for large datasets. The proposed framework explicitly introduces a preconditioning step using an incremental proper orthogonal decomposition to DMD and mode selection algorithms. By performing the preconditioning step, the DMD and the mode selection can be performed with low memory consumption and small computational complexity and can be applied to large datasets. In addition, a simple mode selection algorithm based on a greedy method is proposed. The proposed framework is applied to the analysis of a three-dimensional flows around a circular cylinder.
\end{abstract}

Dynamic mode decomposition\cite{Schmid2008,Schmid2010} (DMD) has been often
used to extract important spatial and temporal structures from fluid flow data
since the method was first proposed in 2008.\cite{Schmid2008}
DMD extracts latent dynamic behavior from input datasets by determining the linear dynamical system that best fits the input datasets.
The notable feature of DMD is that each DMD mode has information on its temporal variation, that is, a growth rate and oscillation frequency.
This is similar to global linear stability analysis.\cite{Theofilis2011, Ohmichi2017caf}
However, unlike global linear stability analysis, DMD does not require governing equations that generate the input datasets.
Therefore, DMD can be applied to cases in which the governing equation is very complicated or unknown. 

However, the standard DMD algorithm\cite{Schmid2010} requires a substantial amount of memory when the algorithm is applied to large datasets because the standard DMD algorithm stores all the input datasets simultaneously.
Numerical and experimental datasets for fluid dynamics research are often too large for the application of DMD.
This difficulty can be avoided using incremental DMD algorithms.\cite{Hemati2014,Ohmichi2017}
Incremental DMD incrementally updates the matrices that are a low-dimensional representation of the input datasets each time new data are obtained instead of using all the input datasets simultaneously.
Therefore, incremental DMD does not require a substantial amount of memory and can be applied to large datasets.

Another issue of standard DMD is that it is not easy to select physically important modes from the obtained modes.\cite{Taira2017}
To overcome this difficulty, several algorithms, such as optimized DMD\cite{Chen2012} and sparsity-promoting DMD\cite{Jovanovic2014} have been proposed.
These methods can determine the small number of DMD modes that are able to represent the input datasets with fewer errors.
However, these methods also require a substantial amount of memory when applied to large datasets.

This letter proposes a simple framework of DMD analysis that can be applied to large datasets.
The framework includes proper orthogonal decomposition\cite{lumley1967, Holmes1996} (POD), DMD, and the mode selection of DMD modes. 
In the proposed framework, preprocessing steps are introduced before performing DMD and mode selection.
Specifically, a small number of principal components of the input datasets are extracted using an incremental-type POD algorithm, then the input datasets are projected onto the principal component bases (namely, POD bases) to reduce the dimension.
Subsequently, the DMD and mode selection steps are performed using the low-dimensionalized datasets.
Mode selection is performed using a simple algorithm based on the greedy approach.
The proposed framework is applied to a three-dimensional flow field around a circular cylinder to demonstrate its effectiveness.

First, as the preconditioning step, the low-dimensionalization of the input datasets is performed using POD.
We use incremental-type POD algorithms because standard POD algorithms also need to store all the input datasets in memory.
There are many incremental-type POD algorithms in the literature (e.g., Refs\cite{Arora2012, Weng2003}).
In this letter, the incremental POD proposed by Arora et al.\cite{Arora2012} is used.
It has been shown that the incremental POD is an effective tool for extracting dominant structures from fluid flow datasets.\cite{Ohmichi2017}

We denote the input datasets as $X = \left[ \bm x_1~\bm x_2~\cdots~\bm x_N\right]$, where a column vector $\bm x_n \in \mathbb{R}^d$ represents the input data at time $t=n \Delta t$.
The incremental POD updates the POD bases using the following algorithm each time new data $\bm x_n$ is obtained.
Suppose that $C_{n-1} \in \mathbb{R}^{l \times l}$ is a rank-$l$ approximation of a covariance matrix constructed using the datasets $\left[ \bm x_1~\bm x_2~\cdots~\bm x_{n-1} \right]$, and its eigendecomposition is $C_{n-1}=U_{n-1} D_{n-1} U_{n-1}^T$,
where the orthogonal matrix $U_{n-1} \in \mathbb{R}^{d \times l}$ and diagonal matrix $D_{n-1} \in \mathbb{R}^{l \times l}$ represent the POD bases and corresponding eigenvalues, respectively.
The POD bases are obtained using the eigendecomposition of the approximate covariance matrix $C_{n-1}$.
The update rule for $C_{n-1}$ is as follows:
First, an average $\bm \mu_{n-1}$ can be updated by
\begin{equation}
    \bm \mu_{n} = \frac{n-1}{n} \bm \mu_{n-1} + \frac{1}{n}\bm x_{n}, \label{Eq.mu}
\end{equation}
and then we define
$\hat{\bm{x}}_{n} = U_{n-1}^T \tilde{\bm{x}}_{n}$
and
$\hat{\bm x}_{n}^\perp = \tilde{\bm x}_{n} - U_{n-1} U_{n-1}^T \tilde{\bm x}_{n}$
where
$\tilde{\bm x}_{n} = {\bm x}_{n} - \bm \mu_{n}$.
$\tilde{\bm x}_{n}$, $\hat{\bm x}_{n}$, and $\hat{\bm x}_{n}^\perp$ represent
the fluctuating component of ${\bm x}_{n}$,
and the parallel and orthogonal components of $\tilde{\bm x}_{n}$ with respect to $U_{n-1}$, respectively.
Using these variables, the covariance matrix $C_{n-1}$ can be updated as follows\cite{Arora2012}:
\begin{equation}
    C_{n} =
    \begin{bmatrix}
    	U_{n-1}~~ & \frac{\hat{\bm x}_{n}^\perp}{\|\hat{\bm x}_{n}^\perp\|}
    \end{bmatrix}
    Q_{n}
    \begin{bmatrix}
    	U_{n-1}~~ & \frac{\hat{\bm x}_{n}^\perp}{\|\hat{\bm x}_{n}^\perp\|}
    \end{bmatrix}^T,
\end{equation}
where
\begin{equation}
    Q_{n} = \frac{n-1}{n^2}
    {
    \renewcommand\arraystretch{2}
    \begin{bmatrix}
    	n D_{n-1} + \hat{\bm x}_{n}\hat{\bm x}_{n}^T~~ & \|\hat{\bm x}_{n}^\perp\| \hat{\bm x}_{n} \\
    	\|\hat{\bm x}_{n}^\perp\| \hat{\bm x}_{n}^T~~ & \|\hat{\bm x}_{n}^\perp\| ^2
    \end{bmatrix}
    }.
\end{equation}
Therefore, the updated POD bases $U_{n}$ and corresponding eigenvalues $D_{n}$ are obtained using the eigendecomposition 
$Q_{n} = U'S'U'^T$ and
\begin{equation}
    U_{n} = 
    \begin{bmatrix}
    	U_{n-1}~~ & \frac{\hat{\bm x}_{n}^\perp}{\|\hat{\bm x}_{n}^\perp\|} 
    \end{bmatrix}
    U',~~~D_{n} = S'. \label{Eq.POD}
\end{equation}
If the rank of $C_n$ becomes greater than a user setting parameter $r_p$, we delete the column and row of $U_n$ and $D_n$ that correspond to the smallest eigenvalues so that the rank becomes $r_p$.
The aforementioned update rule is repeated until all the input datasets are used.

The memory consumption of the incremental POD is $O(r_pd)$, whereas that of the standard POD (and DMD) algorithms is $O(Nd)$.
Therefore, we can apply the incremental POD to large datasets by setting the number of POD bases $r_p$ sufficiently small.

Next, we low-dimensionalize the input datasets.
Before that, we perform the orthogonalization of the $r_d~(= r_p+1)$ bases composed of the $r_p$ columns of POD bases $U_N$ and average $\bm \mu_N$ using the modified Gram--Schmidt (MGS) orthogonalization algorithm.
We denote the obtained orthogonal bases as $P$.
Using $P$, the preconditioning (i.e., low-dimensionalization of the input datasets) is achieved using
\begin{equation}
	\tilde X = P^T X, \label{Eq.LowDim}
\end{equation}
where $\tilde X \in \mathbb{R}^{r_d \times N}$ is a small matrix.

Note that this preprocessing step causes a loss of information in the input datasets.
Generally, small spatial structures tend to disappear when low-dimensionalization using POD bases is applied to fluid flow data.\cite{Taira2011a, Taira2011b}
This loss of information decreases as rank $r_p$ increases.
By contrast, the memory consumption and computational complexity of the incremental POD algorithm are $O(r_p d)$ and $O(r_p^2 d)$, respectively.
Therefore, $r_p$ should be determined by considering this trade-off relation.

Most DMD algorithms can be applied to large datasets using the low-dimensionalized input datasets obtained using the aforementioned preconditioning step,
for example, standard DMD\cite{Schmid2010}, total least squares (TLS) DMD (tlsDMD)\cite{Dawson2016, Hemati2015}, and DMD with augmented input data.\cite{Brunton2016, Tu2014}
In this letter, we use tlsDMD proposed by Hemati et al.\cite{Hemati2015} and Dawson et al.\cite{Dawson2016}
Empirically, tlsDMD has good performance in terms of reconstructing input datasets using a small number of DMD modes because tlsDMD can accurately compute the growth rate and frequency of corresponding DMD modes.

DMD algorithms compute the approximate eigenvalues and eigenvectors of the linear operator $A$ that satisfies $X_1 \approx AX_0$, where $X_0 = \left[ \bm x_1~\bm x_2~\cdots~\bm x_{N-1}\right]$ and $X_1 = \left[ \bm x_2~\bm x_3~\cdots~\bm x_N\right]$.
tlsDMD computes the linear operator $A$ using the TLS method, whereas standard DMD\cite{Schmid2010} uses the least squares method.
As a first step of tlsDMD computation, set $r_d$, that is, the number of columns of $P$, so that $r_d < (N-1)/2$ (this is not a strict restriction) and define
$\tilde Z = \left[ \tilde X_0~~\tilde X_1\right]^T = \left[ P^T X_0~~P^T X_1\right]^T$, where $\tilde Z \in \mathbb{R}^{2r_d \times (N-1)}$.
Then, perform the reduced singular value decomposition of
\begin{equation}
	\tilde Z = 
    \begin{bmatrix}
    	\tilde X_0 \\
    	\tilde X_1
    \end{bmatrix}
    =
    U_d \Sigma V^T, \label{tlsDMD1}
\end{equation}
and partition $U_d$ into four $r_d \times r_d$ sub-matrices as
$
U_d = 
    \begin{bmatrix}
    	U_{11} & U_{12} \\
    	U_{21} & U_{22}
    \end{bmatrix}
$.
Note that $U_d$ is a $2r_d \times 2r_d$ matrix because $\Sigma$ has $2r_d$ non-zero singular values.
Using $U_{11}$ and $U_{21}$, the low-dimensional representation of the linear operator $A$ is obtained as
\begin{equation}
	\tilde A = U_{21}U_{11}^{-1}.
\end{equation}
Finally, solve the eigenvalue problem of 
\begin{equation}
    \tilde A \tilde {\bm \phi} = \tilde \lambda \tilde {\bm \phi}, \label{Eq.LowDMD}
\end{equation}
then the approximate eigenvalues of $A$ and the corresponding DMD modes are
obtained as
\begin{equation}
	\lambda = \tilde  \lambda ~~{\rm and}~~ {\bm \phi} = P \tilde {\bm \phi}, \label{Eq.FullDMD}
\end{equation}
respectively.

Another DMD method for large datasets is a streaming DMD proposed by Hemati et al.\cite{Hemati2014}
The main difference between the proposed method and streaming DMD is that in the latter, the updating of POD bases and projection of input datasets onto the bases are performed simultaneously, whereas in the present method, POD is performed as preprocessing for DMD (and mode selection).
Therefore, the streaming DMD is suitable for online processing of streaming data.
The advantage of the proposed framework is that the POD, DMD, and mode selection methods can be considered and performed separately.
Therefore, various POD, DMD, and mode selection methods can be easily applied to this framework, and it is useful for analysis by trial and error.

Finally, algorithms for mode selection and the reconstruction of input datasets using the selected modes are shown.
Selecting physically important DMD modes is important for the understanding of phenomena and constructing reduced order models.
One promising approach to achieve this is to use compressed sensing, which was first introduced to DMD analysis by Jovanovi\'c et al.\cite{Jovanovic2014}
The proposed method in this letter uses the compressed sensing approach.
To adopt compressed sensing, the low-dimensionalized input datasets $\tilde X = P^T X$ created by the preconditioning step are used instead of the raw large input datasets.
Using the eigenvalues $\lambda$ and eigenvectors $\tilde {\bm \phi}$, $\tilde {\bm x}_n$ can be written as the following expression:
\begin{equation}
	\tilde {\bm x}_n = \sum_{i=1}^{r_d} {\alpha_i \tilde {\bm \phi}_i \lambda_i^{n-1}}.
\end{equation}
Therefore, we can approximate $\tilde X$ using the following matrix form: 
\begin{eqnarray}
	\tilde X &\approx& \tilde \Phi D_\alpha V_{\rm and} \\
	         &=& 
	            \begin{bmatrix}
	            	\tilde {\bm \phi}_1 &  \cdots & \tilde {\bm \phi}_{r_d}
	            \end{bmatrix}	
			    \begin{bmatrix}
			    	\alpha_1 &        & \\
			    	         & \ddots & \\
			    	         &        & \alpha_{r_d}
			    \end{bmatrix}
			    \begin{bmatrix}
			    	\lambda_1^0     & \cdots & \lambda_1^{N-1} \\
			    	\vdots          & \ddots & \vdots          \\
			    	\lambda_{r_d}^0 & \cdots & \lambda_{r_d}^{N-1}
			    \end{bmatrix}.\label{approxX}	         
\end{eqnarray}
In Eq. (\ref{approxX}), the diagonal matrix $D_\alpha$ and Vandermonde matrix $V_{\rm and}$ represent the initial amplitudes and temporal variations of the corresponding DMD modes, respectively.
Note that $\tilde \Phi$, $D_\alpha$, and $V_{\rm and}$ are complex-valued matrices. 

Based on the compressed sensing approach, we select physically important DMD modes as the solution of the following optimization problem:
\begin{equation}
	\mathop{\rm minimize}_{\bm \alpha}~~\|{\bm \alpha}\|_0
	~~{\rm subj.~to}~~
	J({\bm \alpha}) \leq \epsilon, \label{L0opt}
\end{equation}
where
\begin{equation}
	J({\bm \alpha}) = \| \tilde X - \tilde \Phi D_\alpha V_{\rm and} \|_2.
\end{equation}
$\| {\bm \alpha} \|_0$ is the number of non-zero elements in ${\bm \alpha} = \left[  \alpha_1~\alpha_2~\cdots~\alpha_{r_d} \right]$, and $\epsilon$ is a small positive number.
Exact solutions of this optimization problem cannot be easily obtained because this problem is a combinatorial optimization problem.
To obtain approximate solutions, Jovanovi\'c et al.\cite{Jovanovic2014} used the L1 regularization $\| {\bm \alpha} \|_1$ instead of the L0 regularization $\| {\bm \alpha} \|_0$.
This approach is the so-called least absolute shrinkage and selection operator\cite{Tibshirani1994} (LASSO).
Algorithms based on the greedy approach are also often used to solve Eq. (\ref{L0opt}).
It is well known that greedy approaches provide good solutions despite their quite simple algorithms.\cite{Natarajan1995}
This letter proposes the method based on the greedy approach.

The proposed method selects the DMD mode (a column of $\tilde \Phi$) that minimizes a residual $J_{\mathcal{S}}$, and adds the corresponding index of the column to a support set $\mathcal{S}$ at each iteration step.
We define the residual as $J_{\mathcal{S}} = J({\bm \alpha}_{\rm sp})$, where ${\bm \alpha}_{\rm sp}$ is a solution of the following optimization problem:
\begin{equation}
	\mathop{\rm minimize}_{\bm \alpha} J({\bm \alpha})~~{\rm subj. to}~{\rm supp}\{{\bm \alpha}\}=\mathcal{S} ,
\end{equation}
where ${\rm supp}\{{\bm \alpha} \}$ is a set of indices of ${\bm \alpha}$ whose elements have non-zero values.
This optimization problem was introduced by Jovanovi\'c et al.\cite{Jovanovic2014} to determine the optimized amplitude of ${\bm \alpha}$ with a fixed sparsity structure, and they found that ${\bm \alpha}_{\rm sp}$ is obtained using the following computation:
\begin{equation}
	{\bm \alpha}_{\rm sp} =
	\begin{bmatrix}
		I & 0
	\end{bmatrix}
	\begin{bmatrix}
		F    & E \\
		E^T  & 0
	\end{bmatrix}^{-1}
	\begin{bmatrix}
		\bm g \\
		0
	\end{bmatrix},
\end{equation}
where $E$ is a matrix composed of unit column vectors for which the positions of the non-zero elements
correspond to the zero components of ${\bm \alpha}$,
for example,
$
E = 
    \begin{bmatrix}
    	0 & 1 & 0 \\
    	0 & 0 & 1
    \end{bmatrix}^T
$
for
$
{\bm \alpha} = 
    \begin{bmatrix}
    	\alpha_1 & 0 & 0
    \end{bmatrix}^T
$.
$F$ and $\bm g$ are $F = (\tilde \Phi^*\tilde \Phi) \circ  (\overline{V_{\rm and}V_{\rm and}^*})$ and $\bm g = \overline{ {\rm diag} (V_{\rm and} \tilde X^T \tilde \Phi)}$, respectively.
An asterisk denotes the conjugate transpose, a overline denotes the complex conjugate,
$\circ$ denotes elementwise multiplication, and ${\rm diag}(\cdot)$ denotes a vector
whose elements are the diagonal elements of the matrix. 
Note that the computational cost to calculate $J_{\mathcal{S}}$ is not expensive because the dimension of $\tilde X$ is typically small (e.g., $r_d = O(10^{1})$ and $N = O(10^{2})$) as a result of the preconditioning step.
If a stopping criterion set by a user is satisfied, the iteration is terminated.
In this letter, $\|{\bm \alpha}\|_0 = K$ is used as the stopping criterion, where $K$ is a user setting parameter and the number of DMD modes to be selected.
The proposed algorithm is summarized in Table \ref{algorithm}.
\begin{table}[tbhp]
  \centering
  \caption{Mode selection algorithm based on greedy method.\label{algorithm} }
  \begin{tabular}{l} \hline\hline
    \bf{Initialize}:\\
    initial support $\mathcal{S} = \emptyset$ \\ \hline
    \bf{Repeat until stopping criterion is met}:\\
    1)Compute $J_{\mathcal{S} \cup \{ j \}}$ for all the column indices $j \notin \mathcal{S}$\\
    where, \\
    ~~~~~~$J_{\mathcal{S} \cup \{ j \}}= \mathop{\rm min}_{\bm \alpha}~J({\bm \alpha})~~
            {\rm subj.~to}~~{\rm supp}\{ {\bm \alpha} \} = \mathcal{S} \cup \{j\}$\\
    2)Add $j_0 = {\rm arg~min}_{j}~J_{\mathcal{S} \cup \{ j \}}$ to the support $\mathcal{S}$\\
    ~~~~~~$\mathcal{S} \gets \mathcal{S} \cup \{j_0\}$
    \\ \hline \hline
  \end{tabular}
\end{table}

Finally, the reconstructed input datasets $X_R$ using the selected modes can be computed as
\begin{equation}
	X_R = P \tilde \Phi D_{\alpha_{\rm sp}} V_{\rm and}. \label{XR}
\end{equation}

The proposed framework for DMD analysis is summarized in Table \ref{framework}.
\begin{table*}[tbhp]
  \centering
  \caption{Proposed framework for DMD analysis of large datasets.\label{framework} }
  \begin{tabular}{l} \hline\hline
    \bf{I.~Preconditioning}:\\
    ~~~a.~Perform Incremental POD using Eqs. (\ref{Eq.mu})--(\ref{Eq.POD})\\
    ~~~b.~Reorthogonalize the POD bases using MGS method\\
    ~~~c.~Low-dimensionalize the input datasets using Eq. (\ref{Eq.LowDim})\\ 
    \bf{II.~Dynamic Mode Decomposition}:\\
    ~~~a.~Perform DMD to the low-dimensionalized datasets using Eqs. (\ref{tlsDMD1})--(\ref{Eq.LowDMD})\\
    ~~~b.~Calculate full-dimension DMD modes using Eq. (\ref{Eq.FullDMD})\\ 
    \bf{III.~Mode selection}:\\
    ~~~a.~Perform mode selection using the low-dimensionalized datasets (Table \ref{algorithm})\\ 
    \bf{IV.~Reconstruction (If needed)}:\\
    ~~~a.~Input datasets can be reconstructed using the selected modes by Eq. (\ref{XR})
    \\ \hline \hline
  \end{tabular}
\end{table*}

To demonstrate the effectiveness of the proposed framework, we applied the framework to the analysis of three-dimensional laminar flow around a circular cylinder.
The Reynolds and Mach numbers based on the diameter $D$ of the cylinder and freestream velocity $U_{\rm ref}$ were $Re = 350$ and $M = 0.2$, respectively.
The numerical simulation was performed using in-house code that has been verified for several analyses.\cite{Ohmichi2016, Ohmichi2017}
The simulation was performed with a sixth-order compact finite difference scheme\cite{Lele1992, Visbal2002} with tenth-order filtering\cite{Gaitonde2000} for spatial discretization, and a third-order TVD Runge--Kutta method\cite{Shu1988} for time marching.
Periodic boundary conditions were applied in the spanwise direction, with the length of the computational domain $L_z = 3D$.
The computational grid was composed of $9.5 \times 10^6$ grid points.
The number of snapshots was $N = 800$, and the time interval for each snapshot was $\Delta t = 0.25$.
Each snapshot $\bm x_n$ was composed of three components of the velocity in the $x$, $y$, and $z$ directions of each grid point. 

\begin{figure}[tbhp]
\centering
\includegraphics{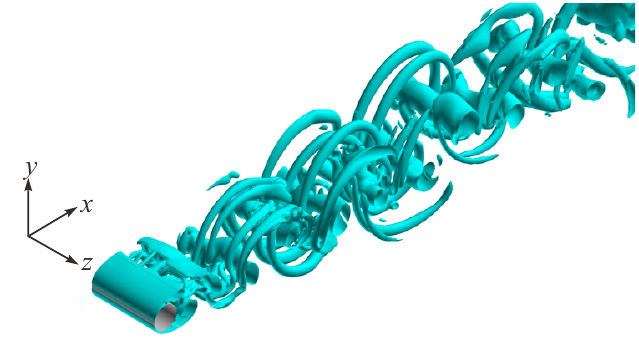}%
\caption{Vortex structures behind the circular cylinder. Iso-surfaces of the $Q$-criterion are shown.\label{isoQ}}%
\end{figure}
Figure \ref{isoQ} shows a snapshot of this flow that extracts vortex structures using the $Q$-criterion.
Although the Reynolds number is relatively low, it is not so easy to extract dominant dynamics from this flow field because the flow includes chaotic behavior.

Memory consumption for the present analysis was approximately 24 GB, with $r_d = 51$.
This memory requirement is sufficiently small for performing the analysis on recent workstations.
Note that if we analyze the present input datasets using standard DMD, over 10 times more memory is required.

\begin{figure}[tbhp]
\centering
\includegraphics{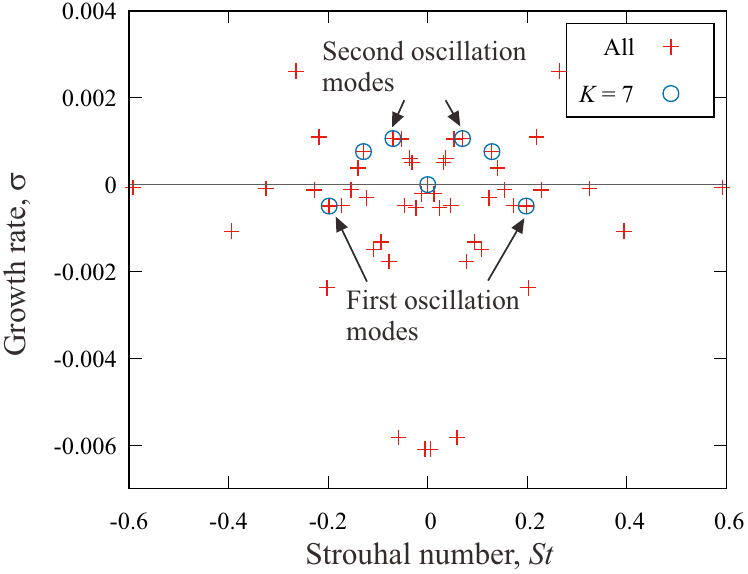}%
\caption{Eigenvalue distribution and seven selected modes. \label{selected_mode}}%
\end{figure}

\begin{figure}[tbhp]
\centering
\includegraphics{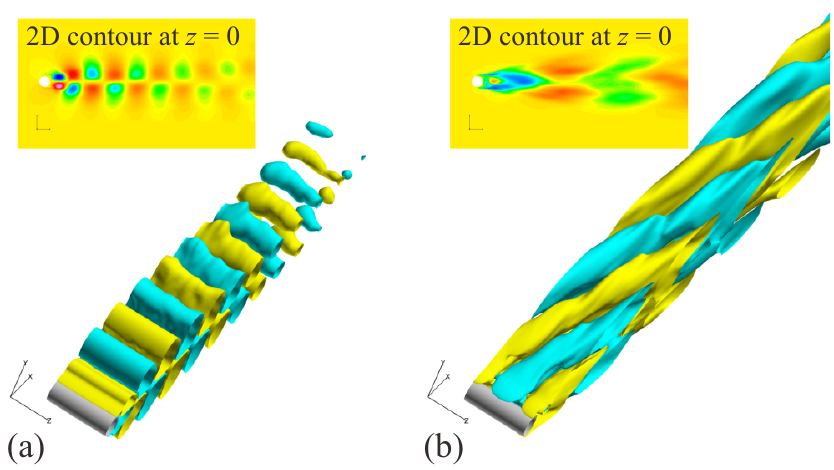}%
\caption{First and second oscillation modes. Iso-surfaces of the velocity in the $x$ direction are shown in yellow and magenta, and denote the opposite phases. (a) First mode of $St = 0.20$. (b) Second mode of $St = 0.07$.\label{DMDmodes}}%
\end{figure}

Figure \ref{selected_mode} shows the distribution of eigenvalues obtained by the present DMD.
The eigenvalues selected by the mode selection algorithm are also shown.
The DMD mode of $(\sigma, St) = (0, 0)$ was first selected, where $\sigma = {\rm Real}\{\rm{log}(\lambda)\} / \Delta t$ and $St = {\rm Imag}\{\rm{log}(\lambda)\} / (2\pi \Delta t)$.
This mode corresponds to the mean flow field.
The first selected oscillation modes had a frequency of $St=0.20$.
This mode corresponds to the well-known two-dimensional vortex shedding, as shown in Figure \ref{DMDmodes}a.
The second oscillation modes had a relatively low frequency of $St = 0.07$.
Interestingly, according to Figure \ref{DMDmodes}b, this mode seems to represent oblique streamwise vortex phenomena.\cite{blackburn2005,jiang2016}
Though not shown here, the third oscillation modes ($St = 0.13$) also represent oblique streamwise vortices.

\begin{figure}[tbhp]
\centering
\includegraphics{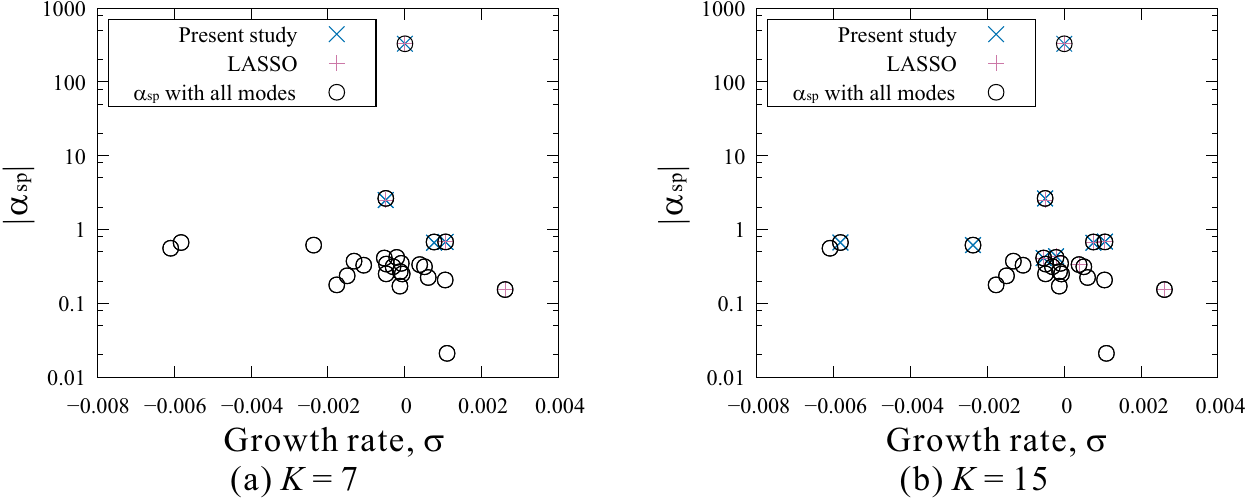}%
\caption{Dependence of the absolute values of the DMD amplitudes $\alpha_{\rm sp}$ on the growth rate $\sigma$. $K$-selected modes obtained by the present mode selection algorithm and LASSO\cite{Jovanovic2014} are shown.\label{alphaVariation}}%
\end{figure}

Figure \ref{alphaVariation} illustrates the dependence of the absolute values of the optimized initial amplitudes $\alpha_{\rm sp}$ on the growth rate $\sigma$.
This figure shows that the modes with small growth rates selected by the mode selection algorithms had large initial amplitudes.
This means that, even if a mode has a small growth rate, the mode can behave as a dominant phenomenon if its initial amplitude is sufficiently large. 
Therefore, both growth rates and initial amplitudes are important in selecting dominant modes that represent the input datasets.
Comparing the proposed mode selection algorithm with the previous algorithm (LASSO)\cite{Jovanovic2014}, it can be seen that the proposed algorithm tends to select modes that have large initial amplitudes rather than large growth rates. 

\begin{figure}[tbhp]
\centering
\includegraphics{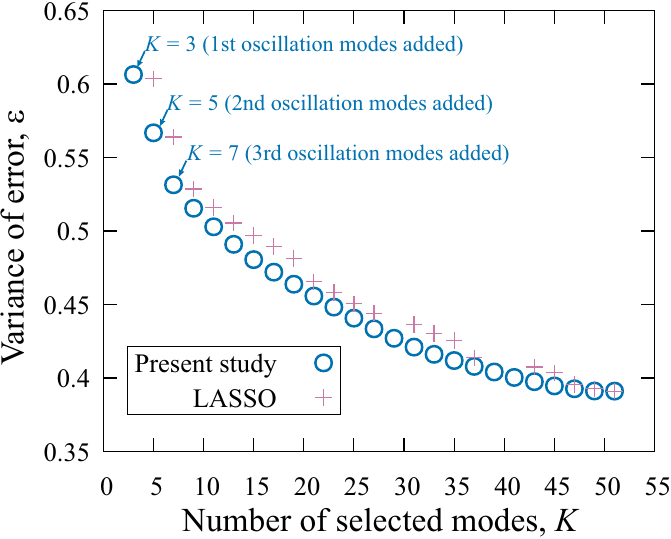}%
\caption{Variance of the reconstruction error against the number of selected modes. \label{variance_of_error}}%
\end{figure}
Figure \ref{variance_of_error} shows the effect of the number of selected modes $K$ used to reconstruct input datasets on the variance of the reconstruction error $\varepsilon$, where $\varepsilon$ is defined by the following equation:
\begin{equation}
	\varepsilon = \frac{ \| X - X_R \|_2^2}{\| \hat X \|_2^2},
\end{equation}
where $\hat X$ is the perturbation component of $X$, that is, $\hat X = \left[ {\bm x}_1 - {\bm \mu}_N~{\bm x}_2 - {\bm \mu}_N\cdots {\bm x}_N - {\bm \mu}_N  \right]$.
Figure \ref{variance_of_error} clearly shows that the error $\varepsilon$ monotonically decreases as $K$ increases.
In particular, the first oscillation modes made a large contribution to the reconstruction of the input datasets.
Additionally, the second and third oscillation modes also made a relatively large contribution.
This means that the proposed mode selection algorithm correctly selected the dominant DMD modes.
Furthermore, in this fluid dataset case, the proposed algorithm had fewer reconstruction error than LASSO.\cite{Jovanovic2014}

\begin{figure}[tbhp]
\centering
\includegraphics{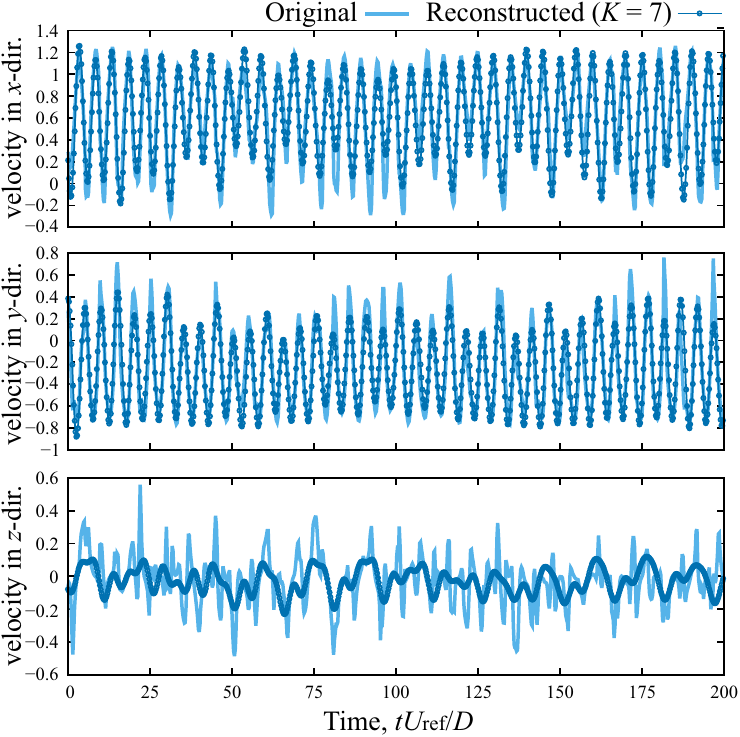}%
\caption{Original and reconstructed ($K=7$) signals at the point $(x, y, z) =(1.51D, 0.51D, 0D)$. \label{reconstructed_signal}}%
\end{figure}

Figure \ref{variance_of_error} indicates there are still certain errors, even if all the modes are used for the reconstruction because of chaotic behavior included in the present flow.
Generally, most DMD algorithms cannot manage the dynamics that cannot be approximated by a local linear operator.
Additionally, the preconditioning step, that is, the dimensionality reduction using POD bases, causes a loss of information, mainly about small spatial structures.
It is necessary to note that the proposed method is suitable for extracting the dynamics of large spatial structures where local linear approximation is valid.
The temporal history of the original and reconstructed flows at the point $(x, y, z) =(1.51D, 0.51D, 0D)$ is shown in Figure \ref{reconstructed_signal}.
We can confirm that large parts of the fluctuations of the velocity are well reproduced using seven modes, although the velocity in the $z$ direction has a relatively large reconstruction error because of its chaotic behavior.

This letter proposed a simple and efficient framework of DMD and its mode selection for large datasets.
The analysis of three-dimensional flow around a circular cylinder showed that the proposed framework can perform DMD and mode selection with low memory consumption, and automatically extracts physically important modes.\newline

The author acknowledges support from JSPS KAKENHI (grant number JP16H01563).
The numerical computations were performed on the JAXA Supercomputer System 2 (JSS2).

\bibliographystyle{unsrt}
\bibliography{reference.bib}

\end{document}